\documentstyle[12pt,titlepage]{article}
\input epsfig.sty

\renewcommand{\theequation}{\thesection.\arabic{equation}}

\font\grande=cmr10 scaled \magstep4
\font\medio=cmr10 scaled \magstep2
\outer\def\beginsection#1\par{\medbreak\bigskip
      \message{#1}\leftline{\bf#1}\nobreak\medskip\vskip-\parskip
      \noindent}

\def\laq{\raise 0.4ex\hbox{$<$}\kern -0.8em\lower 0.62
ex\hbox{$\sim$}}
\def\gaq{\raise 0.4ex\hbox{$>$}\kern -0.7em\lower 0.62
ex\hbox{$\sim$}}

\begin{document}
\bibliographystyle{unsrt}
\titlepage
\begin{center}
{\grande Singularity free dilaton-driven cosmologies}\\ 
\vspace{5mm}
{\grande and pre-little-bangs}\\
\vspace{15mm}
\centerline{Massimo Giovannini \footnote{Electronic address:
m.giovannini@damtp.cam.ac.uk}}
\bigskip
\centerline{{\em DAMTP, 21 Silver Street, Cambridge, CB3 9EW, United
Kingdom}}
\smallskip
\end{center}
\vspace{2cm}
\centerline{\medio  Abstract}
\vspace{10mm}
There are no reasons why the singularity in the growth of
the dilaton coupling should not be regularised, in a string
cosmological context, by the presence of classical inhomogeneities. We
discuss a class of inhomogeneous dilaton-driven models whose
curvature invariants are all  bounded and regular in time and space.
We prove that the non-space-like  geodesics of these models  are all
complete in the sense that none of them reaches infinity for a finite
value of the affine parameter. We conclude that our examples
represent truly singularity-free solutions of the low energy beta functions.
We discuss some symmetries of the obtained solutions and we
clarify their physical interpretation. We also give examples of
solutions with spherical symmetry.
In our scenario each physical quantity is everywhere defined in time
and space, the big-bang singularity is replaced by a maximal
curvature phase where the dilaton kinetic energy reaches its maximum.
The maximal curvature  is always smaller than one (in
string units) and the coupling constant is also  smaller than one and it
grows between two regimes of constant dilaton, implying, together
with the symmetries of the solutions,  that higher genus and higher
curvature corrections  are negligible. We argue that our examples 
describe, in a string cosmological context, the occurrence of 
``little bangs''(i.e.  high curvature phases 
which never develop physical singularities). They also suggest the
possibility of an unexplored ``pre-little-bang'' phase.

\noindent

\newpage
\renewcommand{\theequation}{1.\arabic{equation}}
\setcounter{equation}{0}
\section{Introduction and motivations}

If a consistent approach to the unification of gravitational and gauge
interactions exist at all, it will certainly have important
consequences for cosmology and high energy physics. From a
phenomenological point of view, we would demand, in this framework,
convincing answers to many puzzling questions which arose in
cosmology through the years.
If such a theory exist, it will have to show how the inflationary
paradigm can be  naturally  implemented \cite{1}, how the (general
relativistic) singularities may be regularised \cite{2}, which is the
role played, in the early Universe, by the gauge fields. In one word
which is the theory of the big-bang. A great effort has been done, in
this direction, in the context of superstring inspired cosmological
models. Quite interesting results were obtained. For example it
turned out that (super)inflationary phases with growing curvature and
dilaton coupling are a direct consequence of the classical
solutions of the low energy beta-functions \cite{3,3b}. New
symmetries (like the scale factor duality and the generalised
$O(d,d)$ invariance) appeared to relate the ordinary decelerated
Friedmann-Robertson-Walker expanding phases to their
super-inflationary counterparts \cite{4}. Appealing mechanisms for the
amplification of the vacuum fluctuations were exploited in order to
derive various phenomenological implications of those models
\cite{5,5b}. Of course pre-big-bang models also suffer of various
problems \cite{6}. Among them, it is certainly relevant to mention
the problem of the space-time singularity \cite{7}.

The study of the cosmological solutions of the tree-level low energy
superstring effective action showed the occurrence of different types
of cosmological singularities which seemed indeed quite difficult to
regularise. For example in the context of pre-big-bang models \cite{8}
 it was argued
that regular  solutions of the tree level beta functions were very
hard to find (if not impossible). The
conclusion of those studies was that the only possible way in order
to have regular solutions was to invoke the role played by the
corrections to the tree-level action. In particular it is currently
argued that the corrections arising from the string
tension expansion could regularise the curvature singularities
appearing at tree-level \cite{7} . 

In this paper we are going to explore a different approach to the
singularity problem in the context of string cosmology. Namely we
will assume from the very beginning that the Universe might have been
 inhomogeneous around the singularity and we will try to
understand if an inhomogeneous phase could regularise the solutions
of the low energy beta functions.  In fact the main limitations of
previous numerical studies of string cosmological singularities was
that the Universe was assumed to be completely homogeneous during all
its stages of evolution. In the past the possibility of regularising the 
inhomogeneities through an inhomogeneous phase was indeed discussed
\cite{9}. The attempt was to apply an $O(d,d)$ transformation to 
a (generically singular) $1+1$ dimensional cosmological model
 in order to obtain a higher dimensional (regular) model.
 In that investigation, however, no
examples of inhomogeneous models with growing dilaton coupling
were given and the obtained regular solutions contained an
antisymmetric tensor field and an oscillating dilaton coupling. Moreover
no systematic study of the completeness of the geodesics was
performed. We want also to mention that recently (singular)
inhomogeneous string cosmological models were studied from different
points of view \cite{3,barrow}.

In \cite{m}  it was shown that it is perfectly
possible to find analytical examples of regular inhomogeneous
solutions with growing dilaton coupling. Moreover those solutions
were shown to be regular in the String frame but also in the Einstein
frame \cite{m}. Our point is now very simple.
By itself the regularity of the curvature invariants is not
sufficient in order to state that a  certain solution is free of
singularities. It is well
known that it is quite possible to construct regular solutions whose
geodesics are not complete. A typical example of this behaviour is an
eternally inflating de Sitter model (with scale factor $a(t)= \exp{[H
t]}$) which is clearly not geodetically complete \cite{vilenkin} since
the space-time has an edge. 

The absence of geodesic completeness can indicate two possible
situations. Either the geodesics incompleteness signals a removable
singularity (like in the de Sitter case) 
and then a regular extension of the space-time is
possible, or it shows some true singularity of the space-time. Of
course, if the singularity in the geodesics turns out to be
removable, the question arise of which extension to choose among all
the ones theoretically possible.

In this paper we want to show that it is not only possible to find
regular (inhomogeneous) string  cosmological models without invoking
the role played by the string curvature corrections, but, at the same
time we want to show that the geodesics constructed from the
solutions are all complete in the sense that they can be extended
arbitrarily to any value of the affine parameter.

We find that, before playing with the idea of an inhomogeneous
bangs described on the basis of the low-energy beta functions, it
is better, from a logical point of view, 
 to make sure that the solutions we found are
defined for any value of the affine parameter. If this would not be
the case, any kind of cosmological solution, even the most appealing
one, should be discarded for the simple reason
 that, in spite of the absence of any pole in the curvature
invariants, a boundary in the geodesics defines an unphysical region
of the classical solution.

In the context of general relativity the idea that cosmological  singularities 
could have been regularised via the introduction of strong
 inhomogeneities received, in the past, a lot of attention.
In particular it was shown that it is perfectly possible to find
regular (inhomogeneous) solutions with sources whose energy-momentum tensor 
obey the equation of state of a (perfect) radiation fluid \cite{10}. 
This solution was proven to be geodetically complete \cite{11}, meeting
then all the requirements for a singularity free model. 
Moreover recently these ideas were further explored with various 
extensions aiming 
at more realistic cosmological scenarios \cite{12}.
For earlier discussions of inhomogeneous cosmological models in 
general relativity
one can find useful sources of informations in Ref. \cite{13,14}.
Other useful references (not directly related to our paper but more to
the graceful exit problem in singular string cosmological models) can
be found in Ref. \cite{grace}.

Our paper is organised as follows. In Section II we will review the
main notations and the basic equations involved in the study of
 inhomogeneous string cosmological models. In Section III
we will study the geodesic completeness of our solutions whereas in
Section IV we will contrast the behaviour of our solutions with the
classical singularity theorems. In Section V we will give some
examples of regular spherically symmetric string cosmological
backgrounds admitting a pre-little-bang phase and in Section VI we
will give our concluding remarks.

\renewcommand{\theequation}{2.\arabic{equation}}
\setcounter{equation}{0}
\section{Inhomogeneous dynamics of pre-little-bangs}

There are no 
reasons why a phase of growing dilaton coupling should not be
compatible with a highly inhomogeneous phase sufficiently close 
to the big-bang singularity.
In String theory the maximal curvature scale is
related to the inverse of the String length ($\lambda^{-1}_{s}$) and
the question is if we are allowed to describe such a regime with the
tree-level effective action. The answer to this question is of course
yes, provided the curvature invariants will stay small, in string
units, all along the dynamical evolution. The second physical
requirement one has to demand is that not only the curvature
invariants are bounded but also that the string coupling will stay
much smaller than one . If this would not be the case , we should add
to the effective action also the loop (genus) corrections. In the
context of the pre-big-bang scenario the big-bang was always
associated with the maximal curvature scale of the model. In that
context the dynamics of the bang in rooted in the presence of higher
order curvature corrections. 

In our context the dynamics of the bang
(i.e.  the mechanism able to regularise the curvature) is connected
with the presence of inhomogeneities which allow a completely smooth
evolution between two asymptotic regimes of constant dilaton
coupling. Moreover, recently, the dynamics of the pre-big-bang models was 
analysed using interesting numerical techniques and it has been  argued that 
starting from quite general ``asymptotically trivial'' initial 
conditions certain portions of the parameter space will collapse towards a 
singularity where the string tension corrections will hopefully \cite{7} 
trigger a constant curvature phase. We remind that these studies also foresee 
that other regions of the initial parameter space will not collapse towards
 a singularity. In spite of the fact that in Ref. \cite{3} the spherical 
symmetry was always assumed in order to discuss the past asymptotically 
trivial state according to our picture it is perfectly 
possible that starting from a regime of small curvature and couplings 
some regions will not end up in a singularity but in a perfectly regular 
background whose maximal curvature can be as large as $0.1$ (in string units).
We will name these physical situations ``little bang''. The bang
occurs because the curvature scale reaches a maximal value of the order of 
(but smaller  than) $\lambda_{s}^{-2}$. At the same time the bang is small
if compared to the typical (singular) bangs taking place in the
scenarios of \cite{3}.  In Section V 
we will elaborate more on the role played by
spherical symmetry in the context of little bangs.

Consider the low energy string effective action in the absence of any
dilaton potential
\begin{equation}
S= - \frac{1}{\lambda^2_{s}}
\int d^4 x \sqrt{-g} e^{-\phi} \biggl[ R + g^{\alpha\beta}
\partial_{\alpha}\phi\partial_{\beta} \phi \biggr].
\label{action}
\end{equation}
We deal, more specifically, with the case where the antisymmetric tensor
field has been set to zero.
In this approximation the tree-level beta functions can be obtained
by  direct variation of the action with respect to the metric tensor
and to the dilaton field. After a linear combination of the 
obtained equations we have:
\begin{eqnarray}
& & R - g^{\alpha\beta}\partial_{\alpha}\phi\partial_{\beta}\phi + 2
\Box \phi=0,
\label{beta1}\\
& & R_{\mu}^{\nu} + \nabla_{\mu}\nabla^{\nu} \phi=0.
\label{beta2}
\end{eqnarray}
We consider now the line element of an inhomogeneous cosmological model
with two space-like killing vector fields which admit an Abelian group
$G_2$ on space-like orbits $S_2$ \cite{kramer}
\begin{equation}
ds^2 = A(x,t)[ dt^2 - dx^2] - B(x,t) \biggl[ C(x,t) dy^2 +
\frac{dz^2}{C(x,t)}\biggr].
\label{line}
\end{equation}
Assuming that the space-time dependence of the metric functions can
be factorised as :
\begin{eqnarray}
A(x,t) &=& a(t)~\alpha(x),
\nonumber\\
B(x,t) &=& b(t)~\beta(x),
\nonumber\\
C(x,t) &=& c(t)~\kappa(x),
\label{ansatz}
\end{eqnarray}
 a  solution of the low-energy beta functions can be
\cite{m}, provided,
\begin{eqnarray}
A(x,t) &=& e^{\alpha gd(\mu t)}\biggl[ \cosh{\mu t}\biggr]^{2 +
\beta} \biggl[ \cosh{\biggl(\frac{ \mu x}{2}\biggr)}\biggr]^{ 2 \beta
(\beta +1)},
\nonumber\\
B(x,t) &=& e^{\alpha gd(\mu t)} \cosh{\mu t} \sinh{\mu x},
\nonumber\\
C(x,t) &=& \biggl[ \cosh{\mu t}\biggr]^{1 + \beta} \biggl[
\sinh{\biggl(\frac{\mu x}{2}\biggr)}\biggr] \biggl[
\cosh{\biggl(\frac{\mu x}{2}\biggr)}\biggr]^{ 1 + 2 \beta},
\label{metric}
\end{eqnarray}
and
\begin{equation}
\phi (t) = \alpha gd(\mu t) + \gamma.
\label{dilaton}
\end{equation}
This solution has two obvious symmetries which are also present in the original
action, namely it is invariant under dilaton shift 
($\phi\rightarrow \phi +{\rm constant}$) and coordinate 
reparametrization ($x_{\mu} \rightarrow \mu x_{\mu}$). 

In terms of this symmetry  
the physical interpretation of this solution given in Eqs. (\ref{metric}) 
and (\ref{dilaton})
 is very simple. The
parameter $\mu$ has the dimensions of $[L]^{-1}$ and it fixes, in
string units, the maximal curvature scale accessible. Since we demand
regular solutions below the string scale, we have to require that
$\mu< \lambda^{-1}_{s}$. In the case $\mu > \lambda^{-1}_{s}$ we
always have regular solutions, but, in this regime curvature
corrections will become leading and, therefore, further $\alpha'$
corrections should be added to the (tree-level) effective action
which we used in order to  discuss our results. 

Since $\beta=
\sqrt{\alpha^2 + 4} $ the only other physical parameter of the
solution is exactly $\alpha$.  Again the physical interpretation of
this second parameter is quite simple: it measures, in string units,
the maximal energy of the dilaton background since $\mu^2 \alpha^2 =
\dot{\varphi}^2(0)$ (where the over dot stands for a derivation with
respect to $t$). 

There are also other ``accidental'' symmetries in this solution. 
We find that the metric is left invariant by a parity
transformation, and, therefore given a solution and sending
$x\rightarrow - x$ we obtain again a solution of the low energy beta
functions.
There is also another useful symmetry we want to mention. Namely, if
$\beta \rightarrow -\beta$ and, at the same time, $\alpha\rightarrow
-\alpha$ we get another solution of the low energy beta functions. If
the original solution had growing coupling, the transformed one will
have decreasing coupling. Notice that this symmetry seems to be
connected to the inhomogeneous dynamics of the solution more than with the 
original symmetry of the theory.

An interesting observation is that, since this family of solutions has 
regular curvature invariants for every value of $t$ the only curvature 
singularities may appear for $x=0$. The choice made in the derivation given
in Eqs. (4.2) and (4.4) of Ref. \cite{m} eliminates this possibility.

\renewcommand{\theequation}{3.\arabic{equation}}
\setcounter{equation}{0}
\section{Geodesic Completeness}

In our previous discussion we did not claim that the
solutions given in Eqs. (\ref{metric}) and (\ref{dilaton}) 
were singularity free and we always spoke about ``regular'' solutions.  We
could in principle have manifolds whose curvature invariants are
regular but whose non-space-like geodesics reach a singularity in a
finite proper time. This property of singular manifolds is
generically called {\it geodesic incompleteness} and it means that
the space-time contains non-space-like (i.e. either time-like or null)
geodesics which (when maximally extended) have no endpoint in the
regular part of the manifold.

 If
every non-space-like geodesic can be extended to arbitrary values of
its affine parameter, then we can say that the manifold is {\it
geodetically complete}.
The purpose of the present section is to show that the solution given
in Eqs. (\ref{metric}) and (\ref{dilaton}) is geodetically complete.
To do this we have to show that every non-space-like geodesic has no
end points. In other words we have to show that non-space-like
geodesics can be extended to any arbitrary values of their affine
parameter.

The first step in order to undertake this program is to write down
the geodesics associated with the metric (\ref{metric}), namely
\begin{eqnarray}
&&t'' + \frac{\mu}{2}\Biggl\{\frac{ \alpha + (\beta + 2)\sinh{\mu
t}}{\cosh{\mu t}} \biggr\}\bigl[t'^2  +x'^2\bigr]+ \biggl\{\mu
\beta(\beta + 1) \tanh{\biggl(\frac{\mu x}{2}\biggr)}\biggr\} x' t'
\nonumber\\
&&+ \mu \biggl\{ \frac{\bigl[\sinh{\frac{\mu x}{2}}\bigr]^{2}\bigl[
\alpha + (\beta + 2) \sinh{\mu t}\bigr]}{\bigl[\cosh{\frac{\mu
x}{2}}\bigr]^{2 \beta^2 -2}\cosh{\mu t}}\biggr\}y'^2
+ \mu \biggl\{\frac{ \alpha - \beta \sinh{\mu
t}}{\bigl[\cosh{\bigl(\frac{\mu x}{2}\bigr)}]^{4 \beta + 2 \beta^2 }
\bigl[ \cosh{\mu t}\bigr]^{2 \beta + 3}}\biggr\} z'^2=0,
\label{T}\\
&&x'' + \frac{\mu}{2}\biggl\{ \beta(\beta + 1) \tanh{\biggl(\frac{\mu
x}{2}\biggr)}\biggr\}\bigl[ t'^2 + x'^2\bigr] + \mu\biggl\{\frac{
\alpha + (\beta + 2) \sinh{\mu t} }{\cosh{\mu t}}\biggr\} x' t'
\nonumber\\
&&+\frac{\mu}{4}\biggl\{ \sinh{\mu x}
\frac{\bigl[\beta
- (\beta +2)\cosh{\mu x}\bigr]}{\bigl[\cosh{\bigl(\frac{\mu
x}{2}\bigr)}\bigr]^{2 \beta^2}}\biggr\}y'^2
+ \mu \biggl\{\beta \tanh{\biggl(\frac{\mu x}{2}\biggr)}\frac{ \bigl[
\cosh{\bigl(\frac{\mu x}{2}\bigr)}\bigr]^{-4 \beta
-2\beta^2}}{\bigl[\cosh{\mu t}\bigr]^{ 2 (\beta + 1)}}\biggr\}z'^2=0,
\label{R}\\
&&y'' + \mu\biggr\{ \frac{\bigl[ \alpha + (\beta + 2) \sinh{\mu t}
\bigr]}{\cosh{\mu t}}\biggl\}y' t' + \mu \biggl\{\frac{\bigl[ (\beta
+ 2)\cosh{\mu x} - \beta\bigr]}{\sinh{\mu x}}\biggr\}y' x' =0,
\label{Ph}\\
&& z'' + \mu\biggl\{ \frac{ \bigl[ \alpha - \beta \sinh{\mu
t}\bigr]}{\cosh{\mu t}}\biggr\} z' t' - \mu \biggl\{\beta
\tanh{\biggl(\frac{\mu x}{2}\biggr)}\biggl\} x' z' =0,
\label{Z}\\
&&\bigl[ t'^2  - x'^2 \bigr]
-
2\biggl\{\biggl[\cosh{\biggl(\frac{\mu x}{2}\biggr)}\biggr]^{
2(1-\beta^2)} \biggl[\sinh{\biggl(\frac{\mu x}{2}\biggr)}\biggr]^2
\biggr\}y'^2
\nonumber\\
&&- 2\biggl\{ \biggl[ \cosh{\biggl(\frac{\mu
x}{2}\biggr)}\biggr]^{-2\beta(\beta + 2)} \bigl[\cosh{\mu
t}\bigr]^{-2\beta-2}\biggr\} z'^2
= {\bf s} ~e^{-\alpha gd(\mu t)}\biggl\{\bigl[\cosh{\mu t}\bigr]^{ -
(2 + \beta)}\biggl[\cosh{\frac{\mu x}{2}}\biggr]^{ -2 \beta (\beta +
1)}\biggr\}.
\label{L}
\end{eqnarray}
These equations can be easily derived once the Christoffel symbols
have been computed (see Appendix A for more details).

Notice that we shall be interested in the geodesics of the model 
in the String frame since, from the physical point of view, there
are various reasons to prefer the String frame to the Einstein frame (see
also Section 4 about this point).
 
The previous system of coupled differential equations  can be a bit
simplified by integrating  once Eqs. (\ref{Ph}) and (\ref{Z}). The
result is that Eqs. (\ref{R}), (\ref{T}) and (\ref{L}) are left
unchanged, whereas Eqs. (\ref{Ph}) and (\ref{Z}) become first order
\begin{eqnarray}
&&y' e^{\alpha gd(\mu t)} \bigl[\cosh{\mu t}\bigr]^{\beta + 2}
\bigl[\sinh{\mu x}\bigr]^{\beta+ 2}\biggr[\tanh{\biggl(\frac{\mu
x}{2}\biggr)}\biggr]^{-\beta} = {\bf p}
\label{Ph1}\\
&& z' e^{\alpha gd(\mu t)} \bigl[\cosh{\mu t}\bigr]^{- \beta} \biggl[
\cosh{\biggl(\frac{\mu x}{2}\biggr)}\biggl]^{- 2 \beta} = {\bf q}
\label{Z1}\\
\end{eqnarray}
Notice that ${\bf p}$ and ${\bf q}$ are integration constants whereas
${\bf s}$ is nothing but a parameter which takes values $0$ for null
geodesics and $1$ for time-like geodesics. We will be mostly concerned
with the cases of non-space-like geodesics.
Before starting the analysis of the completeness of the geodesics  we
want to remind a standard result of the theory of ordinary
differential equations \cite{eq}. Consider a generic first order
differential equation of the form
\begin{equation}
\frac{d f}{d v} = {\cal A}(f,v)
\end{equation}
and suppose that ${\cal A}(v)$  is a  continuously differentiable
functions of the variable $v$ defined in some interval ${\cal D}$. Then
the solution of this equation exists in the whole interval where $v$
is defined {\it provided} it can be shown that $|df(v)/dv|$ (i.e.
the norm of the first derivative) is always finite for any finite
value of $v$, which, in our case is nothing but the affine parameter.
The fact that the geodesics are never singular for any finite value
of the affine parameter, will imply  that the geodesics
are complete. Therefore the exercise we intend to do is to study the
space of the solutions of the geodesics equations and to show that
the solutions never show up singularities at finite $\lambda$. We
will study future directed geodesics. Namely we will suppose
 that the initial condition for the coupled system of
differential equations are given at some finite value of $\lambda$
(which is arbitrary) and we will follow the geodesics forward in time.
 Of course one could choose to integrate the
equations backward in time and the analysis turns out to be very
similar. We will just report our  calculations in the case of
geodesics propagating towards the future.
In order to show how to use the argument we just stated, we analyse,
as a warm-up a particularly simple case. Consider, for instance,
time-like geodesics which moving radially (i.e. $y'=0$, $z'=0$,
$x'=0$). In this case we have that Eq. (\ref{T}) gives
\begin{equation}
\frac{d g}{d t} + \frac{\mu}{2} \frac{\alpha + \beta\sinh{\mu
t}}{\cosh{\mu t}} g=0,~~~~g= \frac{d t}{d\lambda}
\end{equation}
Trivial integration gives us that
\begin{equation}
\frac{d t}{d\lambda} = e^{-\frac{\alpha}{2} gd(\mu t)}
\bigl[\cosh{\mu t}\bigr]^{-\frac{\beta + 2}{2}}
\label{pureradial}
\end{equation}
(notice that Eq. (\ref{L}) implies that the geodesics will be along
$r=0$).
It is now clear that since $gd(\mu t)$ is a limited function and
since $\beta = \sqrt{\alpha^2 + 4}> 2$, then we have to conclude that
\begin{equation}
\frac{d t}{d\lambda} < e^{ - \frac{\pi~\alpha}{4}}= {\rm constant}.
\end{equation}
The fact that $t'$ is bounded by a constant implies that, in this
specific case, that radial time-like geodesics are complete.
In the two following subsections we will examine all the possible
cases and we will then conclude that according to the argument we
just discussed that the non-space-like geodesics of this metric are
all complete.

\subsection{Light-like geodesics: {\bf s}=0}

Consider now the case of light-like geodesics with $y'=z'=0$.
Direct integration of Eqs. (\ref{T}) and (\ref{L})
gives
\begin{equation}
x' = {\bf u} e^{-\alpha gd(\mu t)}\biggl[\cosh{\mu t}\biggr]^{-
(\beta + 2)},~~~t'=|x'|
\end{equation}
where ${\bf u}$ is nothing but an integration constant. Remember that
in our solution $\beta=\pm \sqrt{\alpha^2 +4}$. This implies that
either we have  $\beta<-2$ or $\beta >2$. If $\beta>2$ we have that
$x' <\exp{[ -\pi\alpha/2]}$ which proves the completeness of the
geodesics. If $\beta <-2$ we must conclude that since $x'$ is always
finite for any finite value of $\lambda$ the geodesics are also
complete.
A similar situation occurs in the case of light-like geodesics with
$x'=y'=0$, $x=0$ light-like geodesics where direct integration of Eqs.
(\ref{T}) and (\ref{Z}) gives
\begin{eqnarray}
&&z'= {\bf q} e^{- \alpha gd(\mu t)} \bigl[ \cosh{\mu
t}\bigr]^{\beta}
\nonumber\\
&& t'= |{\bf q}|\sqrt{2}e^{- \alpha gd(\mu t)} \bigl[\cosh{\mu
t}\bigr]^{-1}
\end{eqnarray}
In this case $t' <\sqrt{2}|{\bf q}|\exp{[- \pi\alpha/2]}$ for any
value of $\beta$ and $z'$ is also finite for any finite value of
$\lambda$.

\begin{figure}
\begin{center}
\begin{tabular}{|c|}
      \hline
      \hbox{\epsfxsize = 8.5 cm  \epsffile{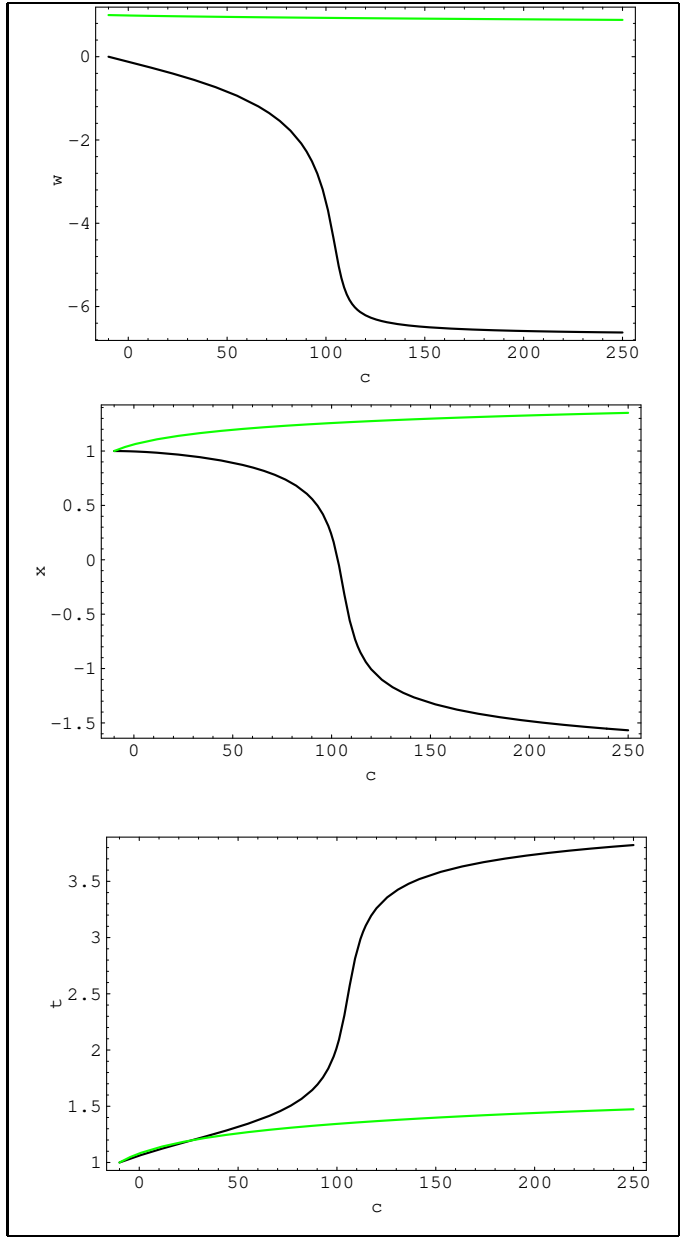}} \\
      \hline
\end{tabular}
\end{center}
\caption[a]{We report some examples of light-like geodesics which are
solutions of the system of differential equations given in Eq.
(\ref{s1}). In this case $\beta =-4$ and we took the initial
conditions to be $x(-10) = 1$, $t(-10)= 1$ and $w(-10) =0$ (full
curves) and $x(-10) =1$, $t(-10) = 1$, $w(-10) =1$ (light curves). On
the horizontal axis is reported the affine parameter $\lambda= c$,
whereas on the vertical axes are reported, in each figure, the
various geodesics and the auxiliary function $w$. We can clearly see
that $t'$ is always positive. If $x'<0$ then we expect that $w'<0$
(since $\beta<-2$). On the other hand, if $x'>0$  we expect that
$w'<0$. In both cases no singularities are present for finite $c$. }
\label{Fig1}
\end{figure}

\begin{figure}
\begin{center}
\begin{tabular}{|c|}
      \hline
      \hbox{\epsfxsize = 8.5 cm  \epsffile{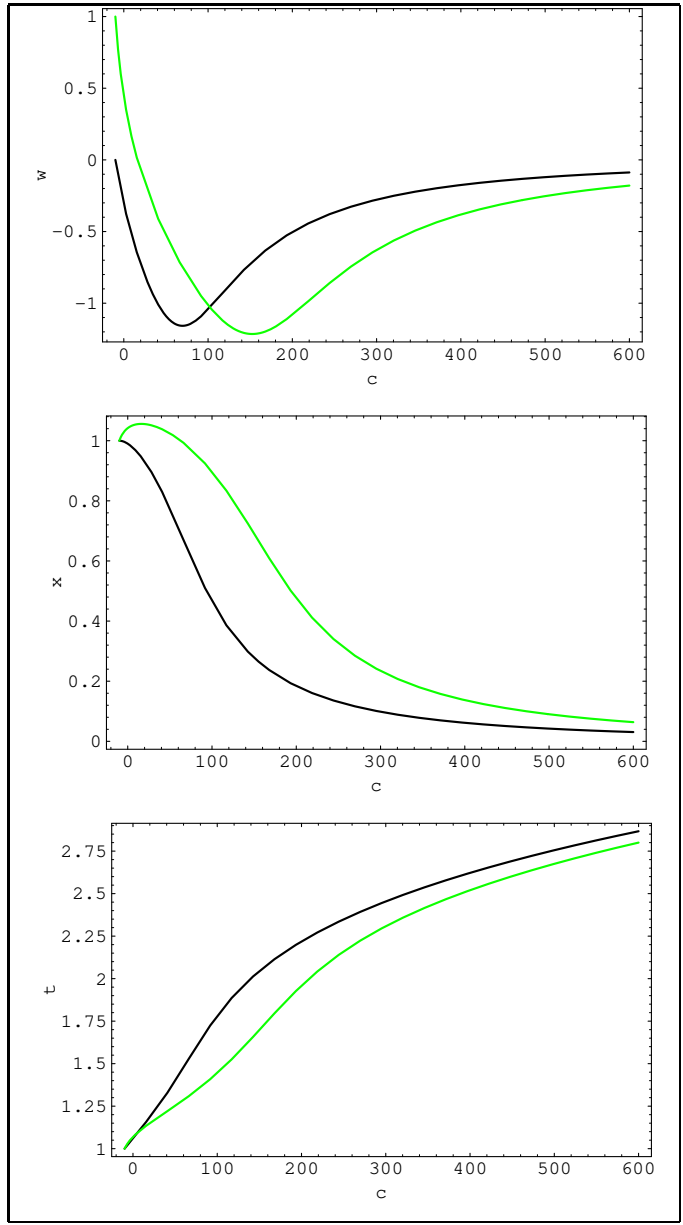}} \\
      \hline
\end{tabular}
\end{center}
\caption[a]{We report the solutions of the system given in Eqs. (\ref{s1})
 with the same initial conditions discussed in the previous figure for 
light and full
 curves, but with $\beta = 4$.}
\label{Fig2}
\end{figure}

We want now to explore the case of light-like geodesics in the case
$y'=0$. In order to do this we go to Eqs. (\ref{T}) and (\ref{L}).
Moreover, from Eq. (\ref{L}) we can see that it is possible to
introduce a new variable which we call $w$. In terms of this new
variable, Eqs. (\ref{L}) and (\ref{Z1}) reduce to
\begin{eqnarray}
&&\frac{d t}{d \lambda} = \cosh{w} {\cal F}(x,t),
\nonumber\\
&&\frac{ d x}{d \lambda} = \sinh{w} {\cal F}(x,t),
\nonumber\\
&&\frac{dw}{d\lambda} = {\cal G}(w,x,t) {\cal F}(x,t),
\label{s1}
\end{eqnarray}
where
\begin{eqnarray}
&&{\cal F}(x,t) =  |{\bf q}| \sqrt{2}e^{ - \alpha gd{\mu t}}
\biggl[\cosh{\biggl(\frac{\mu
x}{2}\biggr)}\biggr]^{-\beta^2}\biggl[\cosh{\mu t}\biggr]^{-1},
\nonumber\\
&&{\cal G}(w,x,t)= - \biggl[ \frac{\mu}{2} \beta (\beta+ 2)
\tanh{\biggl(\frac{\mu x}{2}\biggr)}~ \cosh{w} + \mu (\beta+ 1)
\tanh{\mu t} ~\sinh{w}\biggr]
\end{eqnarray}
The equation for $w'$ was obtained by linearly combining Eqs.
(\ref{T}), (\ref{R}) and (\ref{L}) and by using 
the fact that $t''= x' w'$ (and
that, vice versa,
 $x'' = t' w'$). The evolution for $z'$ is simply given  by Eq.
(\ref{Z1}) and does not need to be re-written. Notice, first of all,
that the sign of ${\cal F}(x,t)$ is always positive for every value
of ${\bf q}$ (which is the integration constant appearing in Eq.
(\ref{Z1})). Suppose that $x'>0$. In this case we have that,  since
$x(\lambda)$ and $t(\lambda)$ are two increasing functions of the
affine parameter, the sign of $w'$ is only determined by the relative
weight of the numerical constants appearing in front of the two
hyperbolic tangents contained in the expression of ${\cal G}(w,x,t)$.
Now, since the curvature invariants are bounded provided $|\beta|>2$
we have that the sign of ${\cal G}(w,x,t)$ is always
negative, and that, therefore, $w(\lambda)$  increases without being
singular for any finite $\lambda$. Thus $w(\lambda)$ does not diverge
for finite $\lambda$ and since $x'$ cannot grow arbitrarily large we
have to conclude that the geodesics are all complete. Suppose now
that $x'<0$. This time we have that, since $\tanh{(\mu x/2)}$  always
decreases as a function of $\lambda$, the sign of $w'$ will be
determined by the sign of $-\mu(\beta + 1)$ which is positive (for
large $\lambda$)if $\beta <-2$ and negative for $\beta >2$. In both
cases $w'$ is always finite. It is also easy to see that this system
of equations does not yield singularities when the geodesics approach
$x$ ($x'<0$, $\sinh{w}<0$). The reason is that $w'$ becomes positive
and prevents the $x$ coordinate from decreasing too quickly. The same
argument can be used in the case of the $t$ coordinate. If $t$ would
grow too fast this would imply  that $x$ would grow or decrease too
quickly. But then the action of $w'$ will again prevent possible
divergences. In Fig. \ref{Fig1} and Fig. \ref{Fig2} we report some
plots of the geodesics studied in this last case (Eqs. \ref{s1}). We
reports the plots only for illustration and in order to corroborate
with some picture the analytical arguments presented up to now. 

\begin{figure}
\begin{center}
\begin{tabular}{|c|}
      \hline
      \hbox{\epsfxsize = 7.5 cm  \epsffile{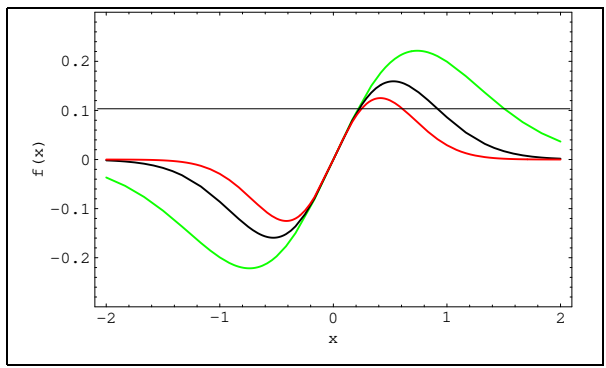}}\\
      \hline
\end{tabular}
\end{center}
\caption[a]{We plot the function $f(x)$ given
in Eq. (\ref{int2}). The three curves from top to bottom correspond,
respectively,  to the cases $\beta =3,~4, ~5$. Notice that in Eq.
(\ref{int1}) the dependence on $\beta$ only occurs through the factor
$(1 -\beta^2)$ and it is, therefore, invariant for $\beta \rightarrow
- \beta$. With the horizontal line is denoted a particular value of
${\bf k}$.}
\label{Fig3}
\end{figure}

We can easily integrate the geodesics also in the
case where $z'=0$. We will have:
\begin{eqnarray}
&&\frac{d t}{d \lambda} = \cosh{w} {\cal M}(x, t)
\nonumber\\
&&\frac{d x}{d \lambda}  = \sinh{w} {\cal M}(x,t)
\nonumber\\
&&\frac{dw}{d\lambda}={\cal M}(x,t) {\cal N}(w,x,t)
\label{sys3}
\end{eqnarray}
where
\begin{eqnarray}
&&{\cal M}(x,t) ={\bf p} \sqrt{2}e^{- \alpha gd(\mu t)}\frac{
\biggl[\cosh{\biggl(\frac{\mu x}{2}\biggr)}\biggr]^{ 1 -\beta-
\beta^2} \biggl[\sinh{\biggl(\frac{\mu x}{2}\biggr)}\biggr]^{1 +
\beta}}{\bigl[\sinh{\mu x}\bigr]^{(\beta + 2)} \bigl[\cosh{\mu
t}\bigr]^{(\beta + 2)}}
\nonumber\\
&&{\cal N}(w,x,t) = -\frac{\mu}{2} \cosh{w} \biggl\{ (\beta^2 - 1)
\tanh{\biggl(\frac{\mu x}{2}\biggr)} - \biggl[\tanh{\biggl(\frac{\mu
x}{2}\biggr)}\biggr]^{-1}\biggr\}
\label{MN}
\end{eqnarray}
In this case the situation seems a little bit more complicated.
Consider, in particular the analytic expression of ${\cal M}(x,t)$.
We can immediately see that for $x\rightarrow 0$, ${\cal M}(x,t)$
might diverge as soon as $x=0$ is reached. Therefore we have first of
all understand if $x=0$ can be ever approached by the geodesics. In
order to do this we can divide $x'$ by $w'$ and find an analytical
relation between $w$ and $x$. The differential relation is:
\begin{equation}
\frac{dx}{d w} = -\frac{2}{\mu} \frac{\tanh{w}}{ (\beta^2 -1
)\tanh{\biggl(\frac{\mu x}{2}\biggr)} - \biggl[\tanh{\biggl(\frac{\mu
x}{2}\biggr)}\biggr]^{-1}}.
\end{equation}
By integrating this equation once we find that
\begin{equation}
\cosh{w} = {\bf k}^{-1} \sinh{\biggl(\frac{\mu x}{2}\biggr)}
\biggl[\cosh{\biggl(\frac{\mu x}{2}\biggr)}\biggr]^{1 - \beta^2}.
\label{int1}
\end{equation}
Notice that ${\bf k}$ is a positive integration constant.
{}From this expression we learn that the solution of the system given
in Eqs. (\ref{sys3}) will never pass through $x=0$. In fact we can
notice that ${\bf k}\cosh{w} \geq {\bf k}$. But if $x=0$, then, the
right hand side of Eq. (\ref{int1}) will also go to zero and this
will mean that ${\bf k}\cosh{w}$ has to go to zero which is never the
case. We can visualise this fact by plotting the function
\begin{equation}
f(x) =\sinh{\biggl(\frac{\mu x}{2}\biggr)}
\biggl[\cosh{\biggl(\frac{\mu x}{2}\biggr)}\biggr]^{1 - \beta^2}.
\label{int2}
\end{equation}
 In Fig. \ref{Fig3} we report this plot. We notice that for each
particular value of ${\bf k}$ there will be two intersection with the
curve $f(x)$ (corresponding to the case $w=0$). This implies that $x$
is always positive and its variation will be bounded between these
two intersections. This implies that ${\cal M}(x,t)$ will also be
positive definite. Since according to the above relation (which has to
hold for any $\lambda$) $\cosh{w}$ will never be zero (by its
definition), $x(\lambda)$ will never reach zero and 
$t(\lambda)$ will be and increasing
function of the affine parameter which never gets singular for any
finite value of $\lambda$. Moreover the $x$ coordinate can only take
values between $x_{+}$ and $x_{-}$ where $x_{\pm}$ are defined from 
the solution of the trascendent equation $f(x_{\pm}) = {\bf k}$.
 The same argument shows that $x'$ is also
finite for any finite value of $\lambda$ and this implies that the
geodesics are complete.

\subsection{Time-like geodesics: {\bf s} =1}

Along the lines of this last case we can treat the case where $y'=0$,
$z'=0$ and ${\bf s}= 1$. In this case it is again useful to introduce
the variable $w$. From Eqs. (\ref{L}) and (\ref{Ph1}) we can see that
\begin{eqnarray}
&&\frac{ d t}{ d \lambda} = \cosh{w} ~ {\cal P} (x,t)
\nonumber\\
&&\frac{ d x}{d \lambda}  = \sinh{w} ~{\cal P}(x,t)
\nonumber\\
&&\frac{d w}{d \lambda} = {\cal R}(w,x,t) {\cal P}(x,t)
\label{p}
\end{eqnarray}
where
\begin{eqnarray}
&&{\cal P}(x,t) = \sqrt{{\bf s}} e^{ -\frac{\alpha}{2} gd{(\mu t)}}
\biggl[\cosh{\mu t} \biggr]^{-\frac{\beta + 2}{2}}\biggl[
\cosh{\biggl(\frac{\mu x}{2}\biggr)}\biggr]^{- \beta (\beta +1)}
\nonumber\\
&&{\cal R}(w,x,t) = -\frac{\mu}{2} \biggl[ \frac{\alpha + (\beta +
2)\sinh{\mu t}}{\cosh{\mu t}} \sinh{w} + \beta(\beta
+1)\tanh{\biggl(\frac{\mu x}{2}\biggr)}\cosh{w}\biggr]
\label{pr}
\end{eqnarray}
From Eqs. (\ref{p}) and (\ref{pr}) we can see that $t'$ is always positive 
for time-like geodesics and it is also finite for any $\beta$. Suppose 
now that $x'>0$. Then the sign of $w'$ will be mainly determined 
by the sign of ${\cal R}(w,x,t)$. Since $x(\lambda)$ and $t(\lambda)$
always increasing functions of the affine parameter we have that 
$w(\lambda)$ increases, without being singular, for any finite $\lambda$. 
Then the geodesics are all complete. With a similar argument we can prove the
 completeness in the case $x'<0$.

If ${\bf s}=1$ we have also to study the case $x'=0$, $y'=0$, $x=0$.
In this case from Eq. (\ref{Z}) we have that the evolution equations
for (\ref{Z}) and (\ref{T}) can be written as
\begin{eqnarray}
&&\frac{d z}{d \lambda} = {\bf q} e^{-\alpha gd(\mu
t)}\biggl[\cosh{\mu t}\biggr]^{\beta} ,
\nonumber\\
&&\frac{d t}{d \lambda} = \biggl[\cosh{\mu t}\biggr]^{-1} \biggl\{ 2
{\bf q}^2 e^{- 2\alpha gd(\mu t)}  + {\bf s} e^{ - \alpha gd(\mu
t)}\biggl[ \cosh{\mu t}\biggr]^{- \beta} \biggl\}^{\frac{1}{2}}.
\end{eqnarray}
Again in this case we see that $z'$ and $t'$ are always finite and no
singularity is encountered for any finite $\lambda$. Moreover,
suppose that $\beta >2$. Then $t'$ is bounded by a constant, namely
\begin{equation}
\sqrt{({2 {\bf q}^2\exp{[ -\pi\alpha]} + {\bf s}\exp{[-\pi\alpha/2]}})}
\end{equation}
while $z'$ is always finite. If, on the other hand, $\beta <-2$ we
see that $z'< |{\bf q}| \exp{[-\pi\alpha/2]}$ while $t'$ is finite
for finite $\lambda$.
\subsection{Non-space-like geodesics}
At this point we are in condition of writing a general form of the
geodesics in the non-space-like case.
\begin{eqnarray}
&&\frac{d t}{d \lambda} = \cosh{w}~ {\cal F}(x,t) {\cal L}(x,t)
\nonumber\\
&&\frac{dx}{d\lambda} = \sinh{w} ~{\cal F}(x,t) {\cal L}(x,t)
\nonumber\\
&&{\cal L}(x,t) = \frac{1}{|{\bf q}| \sqrt{2}}\Biggl\{ 2 {\bf q}^2 +
2 {\bf p}^2 \frac{\biggl[\sinh{\biggl(\frac{\mu
x}{2}\biggl)}\biggr]^{2(\beta + 1)}\biggl[\cosh{\biggl(\frac{\mu
x}{2}\biggr)}\biggr]^{2 (1-\beta)}}{\bigr[\sinh{(\mu
x)}\bigl]^{(\beta + 2)} \bigl[\cosh{(\mu t)}\bigr]^{2(\beta + 1)}}
\nonumber\\
&&+ \frac{{\bf s} e^{\alpha gd(\mu t)}}{ \bigl[ \cosh{\mu
t}\bigr]^{\beta} \biggl[\cosh{\biggl(\frac{\mu
x}{2}\biggr)}\biggr]^{2 \beta}}\Biggr\}
\end{eqnarray}
The evolution of $w$ in terms of the affine parameter can be obtained
after some manipulations.
Define  the functions
\begin{eqnarray}
&&\Lambda_{1}(x,t) = e^{- 2 \alpha gd(\mu t)}\Biggl\{{\bf p}^2 \frac{
\biggl[ \sinh{\biggl(\frac{\mu x}{2}\biggl)}\biggr]^{2(1 + \beta)}
\biggl[\cosh{\biggl(\frac{\mu x}{2}\biggr)}\biggr]^{2( 1  -\beta^2
-\beta)}}{ \bigl[\cosh{\mu t}\bigr]^{(2\beta + 5)} \bigl[ \sinh{\mu
x}\bigr]^{ 2 (\beta + 2)}}
\nonumber\\
&&+{\bf q}^2 \frac{ \alpha  - \beta \sinh{ \mu t} }{\biggl[
\cosh{\biggl(\frac{\mu x}{2}\biggr)}\biggr]^{ 2 \beta^2 } \bigl[
\cosh{\mu t}\bigr]^3}\Biggr\}
\nonumber\\
&&\Lambda_{2}(x,t)  = e^{- 2 \alpha gd(\mu t)}\Biggl\{ \frac{{\bf
p}^2 }{4}\frac{\biggl[\sinh{\biggl(\frac{\mu x}{2}\biggr)}\biggr]^{2
\beta}\bigl[\beta -(\beta + 2) \cosh{\mu x}\bigr]}{\biggl[
\cosh{\biggl(\frac{\mu x}{2}\biggr)}\biggr]^{2 \beta^2 + 2\beta}
\biggl[\sinh{\mu x}\biggr]^{2 \beta + 3} \biggl[\cosh{\mu
t}\biggr]^{2 \beta + 4}}
\nonumber\\
&&+\beta {\bf q}^2\frac{\sinh{\biggl(\frac{\mu x}{2}\biggr)}}{\biggl[
\cosh{\biggl(\frac{\mu x}{2}\biggr)}\biggr]^{2\beta^2 +1}
\bigl[\cosh{\mu t}\bigr]^{2}}\Biggr\}
\end{eqnarray}
then the equation for $w$ becomes
\begin{eqnarray}
\frac{dw}{d\lambda}&=& \cosh{w} ~{\cal F}(x,t) {\cal L}(x,t) \Biggl[
- \frac{\mu}{2} \beta (\beta + 1) \tanh{\biggl(\frac{\mu
x}{2}\biggr)} - \frac{\mu\Lambda_{2}(x,t)}{\bigl[{\cal F}(x,t) {\cal
L}(x,t)\bigr]}\Biggr]
\nonumber\\
&& +\sinh{w} {\cal F}(x, t) {\cal L}(x,t) \Biggl[ - \frac{\mu }{2}
\frac{\alpha + (\beta + 2) \sinh{\mu t}}{\cosh{\mu t}} +
\frac{\mu\Lambda_{1}(x,t)}{\bigl[{\cal F}(x,t) {\cal
L}(x,t)\bigr]}\Biggr]
\label{tot}
\end{eqnarray}
{}From these equations we see that the only possibly dangerous terms
are the ones containing some hyperbolic sinus in the denominator.
These terms could explode for $x\rightarrow 0$. We can see that this
is not the case. Suppose we isolate the possible dangerous
contribution by keeping only the leading terms in the
limit$x\rightarrow 0$.
We see that
\begin{equation}
\lim_{x\rightarrow 0} {\cal F}(x,t) {\cal L}(x,t) = {\cal M}(x,t)
\end{equation}
where ${\cal M}(x,t) $  is simply given by Eq. (\ref{MN}). Recall now
that by expressing $x''$ as $w't'$ and $t''$ as $w't'$, from Eqs.
(\ref{L}) and (\ref{T}) we get an identity which allows to simplify,
in the limit $x\rightarrow 0$, Eq. (\ref{tot}). The final result for
Eq. (\ref{tot}) reproduces exactly Eqs. (\ref{sys3})--(\ref{MN}).
Therefore, the only possible singularities arising in the general
case are the ones already appearing in the case of light-like
geodesics with $z'=0$.  We can repeat the same argument discussed in
that context and conclude that the geodesics are complete also in
this case, since the dangerous value (i.e. $x=0$) cannot be reached
for finite value of $\lambda$. In different limits Eqs. (\ref{tot})
reproduce the various geodesics specifically studied in the previous
subsections. We can say that the general case does not introduce
unexpected singularities and we will finally conclude that the family
of non-space-like geodesics is all complete. The derivatives of each
coordinate are always finite for finite values of the affine
parameter.

From the conclusions of the present section we see that the
solution of the low-energy beta functions discussed in \cite{m} and
illustrated in Eq. (\ref{metric}) represent a class of singularity
free backgrounds. In fact in Ref. \cite{m} the regularity of the
curvature invariants was proven by direct calculation of the scalar
curvature and of the squares of all the other curvature invariants.
In this paper we showed that all the non-space-like geodesics are
complete. Given the (rather surprising) properties of these solutions
of the effective action (\ref{action}) 
we find useful to discuss in a more detailed fashion
the interplay among our geometry and the singularity theorems.

\renewcommand{\theequation}{4.\arabic{equation}}
\setcounter{equation}{0}
\section{Singularity Theorems and  dilaton-driven cosmologies}

From the results discussed in the previous section we can be
confident that our space-time is geodetically complete and that, at
the same time, the curvature invariants are everywhere defined in
time and space without hitting any curvature singularity \cite{m}.

Thus we can conclude that our solutions of Eqs. (\ref{metric}) and 
(\ref{dilaton}) are singularity free. 

Many
questions can arise at this point. For example one could ask which
hypotheses of the singularity theorems turn out to be violated, which
are the asymptotic of our solutions and so on and so forth. 
Before
addressing these questions we want to make a general statement. In
the context of superstring inspired  cosmological models it is quite
important, from the physical point of view, to discuss the
singularity properties of a model both in the String and in the
Einstein frame.
 In the String frame the Planck length is not a
constant and it evolves according to the evolution of the dilaton
coupling. The true constant in the string frame is then  the string
length ($\lambda_{s}$).Indeed in Ref. \cite{m} it was shown that, for
the class of models at hand, the regularity of the curvature
invariants occurs both in the Einstein and in the String frame.

There are physical reasons to prefer the String to the Einstein
frame. For example the study of the motion of classical strings in
cosmological backgrounds showed that the centre of mass of the string
follows geodesics with respect to the String frame \cite{dvs,ng} but not
with respect to the Einstein frame. Moreover, looking at the
properties of the cosmological solutions of the low energy beta
functions it is well known that the super-inflationary expansion
appearing in the String frame description of the pre-big-bang
scenario, turns into an accelerated contraction in the Einstein
frame.

 In this paper we discussed the completeness of the geodesics
in the String frame. It is not true, in general, that the geodesic
completeness in the String frame implies the geodesics completeness
in the Einstein frame. In our case using the fact that the dilaton
coupling (appearing in the conformal factor) is completely regular
and using also the fact that our metric is diagonal \cite{zia}the
geodesic completeness can be shown to hold also in the Einstein frame
and an explicit calculation (which e do not report) tells us that
this is indeed the case.

In more general terms,  the situation is
not so obvious and the singularity properties  of the solutions of
the low energy beta functions cannot be simply translated from the
String to the Einstein frame.
 For example it is almost intuitive that singular metrics in one
frame can become regular in the other. Typically this can happen if
the dilaton is evolving towards a phase of linear growth (in cosmic
time). If this is the case, in spite of the regularity of the
curvature invariants (say in the String frame), it is almost obvious
that the singularity will reappear in the Einstein frame thanks to
the conformal transformation involving the dilaton coupling. This is
exactly what happens in most the regular solutions of the low energy
beta functions \cite{7} which are indeed only regular in the String
frame.
In the terminology of  the present paper we call singularity free
only those solutions which are regular and geodetically complete in
the String frame and in all the other frames conformally related to
it.
We want to recall  that in our examples the dilaton always evolve
between two constant regimes, and, therefore, the regularity in one
frame implies the regularity in the other.

Therefore, a well defined problem would be to precisely formulate the
singularity theorems directly in the String frame. We will not
address this quite important problem in general (even if it would be
by itself of extreme interest). Following the approach adopted in the
previous sections we will only be concerned with our solutions and we
will postpone to future studies the considerations related to more
general metrics (where for instance the dilaton is not going to
constant value but either it oscillates or it increases linearly in
cosmic time).

In this spirit a legitimate question to ask is, in our context if, given a
non-space-like vector $u^{\mu}$, the condition
\begin{equation}
u^{\mu} u^{\nu}  R_{\mu\nu} >0
\label{condition}
\end{equation}
is satisfied or not. 
In the context of the singularity theorems one
often refers to the strong energy condition ($u^{\mu} u^{\nu}
R_{\mu\nu} \geq 0$ for all causal vectors $u^{\mu}$) and to the null
convergence condition ($v^{\mu} v^{\nu}R_{\mu\nu}\geq 0$) for all
null vectors $v^{\mu}$. We will try to avoid using the weak  energy
condition ($T_{\mu \nu} u^{\mu} u^{\nu} \geq 0$ for all causal
vectors $u^{\mu}$)  and the dominant energy condition since they
involve directly the energy-momentum tensor $T_{\mu\nu}$.

The singularity theorems usually invoke an appropriate boundary or
initial condition which tries to express the fact that some finite
portion of the space is trapped within itself and nothing can escape
from it. Since historically the singularity theorems were firstly
applied  to the problem  of the gravitational collapse \cite{pen},
this condition is quite easy to understand physically. In a
cosmological context this condition might look at first sight more
difficult to understand or less motivated. Of course this is not the
case, but, none the less the initial boundary condition is an
absolute necessity for the proof of the singularity theorems in their
strongest version (for a more detailed discussion of this point see
the excellent review of Ref. \cite{rev}).

In more formal terms a trapped surface is a space-like surface whose
two (future directed) null fundamental forms have the same sign. If
both traces are negative the surface is said to be future trapped,
whereas, if they are both positive they are said to be past trapped.

In  our solution something quite amusing happens. In
fact it turns out that some of the models included in the class 
Eqs. (\ref{metric}) and (\ref{dilaton}) violate the condition given in Eq.
(\ref{condition}). Some other models do not violate any causality or
energy condition but do violate the initial boundary condition in the
sense that closed surfaces are not necessarily trapped.

To show this consider the explicit calculation reported in Appendix
B. We see that the condition (\ref{condition}) is not satisfied since
the contraction of the Ricci tensor with two generic non-space-like
vectors changes signs. However, if we choose $\beta =  -4$ the
condition (\ref{condition}) is satisfied. In fact for $\beta =-4$ the
hyperbolic sinuses responsible for the sign flips In Eq.
(\ref{condition}) disappear from the final expression.

If $\beta =-4$ we can show that in spite of the fact that the
condition of Eq. (\ref{condition}) is satisfied,  the space-time does
 contain  closed surfaces that are not trapped. 
The standard technique in order to show
that trapped surfaces are not present is to compute the two null
(future directed) second fundamental forms which are nothing but the
second fundamental forms corresponding to the two (future directed
normals). They are defined as:
\begin{equation}
K^{\pm}_{a b} = e^{\mu}_{b} e^{\nu}_{a} \nabla_{\nu} k^{\pm}_{\mu}
\end{equation}
where $a,b...=2,3$ whereas $\mu,\nu...=0,1,2,3$;
$\vec{e}_{a}=e^{\mu}\frac{\partial }{\partial x^{\mu}}(S) $ denote
the two vectors defined only on the closed surface $S$; $\gamma_{a
b}$ is the (two dimensional) metric on $S$; $k^{\pm}_{\mu}$ are the
two null normal one-forms defined as $k^{\pm}_{\mu}e^{\mu}_{a}=0$,
$k^{+}_{\mu}k^{+~\mu}=0$, $k^{-}_{\mu}k^{-~\mu}=0$, $k^{+}_{\mu}
k^{-~\mu} = 1$; $\nabla_{\nu} $ is the covariant derivative with
respect to the metric $g_{\mu\nu}$. The traces of the two null
fundamental forms are simply given by
\begin{equation}
K^{+}= \gamma^{a b} K^{+}_{a b},~~~K^{-} = \gamma^{a b} K^{-}_{a b}
\end{equation}
Now, the space-like surface $S$ is trapped if and only if the traces
of the two null second fundamental forms have the same signs. In the
case of our metric we find that  the two traces (at a  point $p$
where $|x|$ reaches its maximum on the closed surface $S$)  are given by
\begin{eqnarray}
&&K^{\pm} = g^{yy}\partial_{y} k_{y}(p) + g^{zz}\partial_{z} k_{z}(p) +
\frac{\mu}{2\sqrt{2}}\frac{e^{-\frac{\alpha}{2} gd(\mu t)
}}{\biggl[\cosh{(\mu t)}\biggr]^{\frac{\beta + 2}{2}}
\biggl[\cosh{\biggl(\frac{\mu x}{2}\biggr)}\biggr]^{\beta(\beta +
1)}} \Biggl\{ \frac{2 \alpha + (\beta + 4) \sinh{\mu t}}{\cosh{\mu
t}}
\nonumber\\
&&\mp \biggl[ (\beta^2 + \beta + 1) \tanh{\biggl(\frac{\mu
x}{2}\biggr)} + \biggl(\tanh{\biggl(\frac{\mu
x}{2}\biggr)}\biggr)^{-1}\biggr] \Biggr\}
\label{traces}
\end{eqnarray}

From the above expression we can see that in the limit $\beta =-4$
the first line of Eq. (\ref{traces})  is positive (since $k_z$ and
$k_{y}$ are positive for outgoing normals). However, the second line
is not positive definite and, therefore, the two null fundamental
forms do not have the same sign. Therefore our space-time admits
closed surfaces which are not trapped. Thus the  initial (boundary)
condition of the singularity theorems is not satisfied and this is
the reason why, also in the $\beta= -4$ case we are able to find non
singular solutions.

We want to mention another interesting
properties of our solutions. If we look at the curvature invariants
we said that they are all bounded. Moreover a quite amusing fact
holds. Namely if we look at the ratio of the Weyl invariant compared
to the Riemann invariant \cite{m} we see that
\begin{equation}
\lim_{x\rightarrow \infty,~t\rightarrow \infty
}\frac{C^{\mu\nu\alpha\beta}
C_{\mu\nu\alpha\beta}}{R^{\mu\nu\alpha\beta}R_{\mu\nu\alpha\beta}} =
{\rm constant}
\end{equation}
This asymptotic behaviour, together with the fact that
$R_{\mu\nu}R^{\mu\nu}$ goes to zero (for $[x,t]\rightarrow\infty$)
faster than $ C^{\mu\nu\alpha\beta} C_{\mu\nu\alpha\beta}$ signals
the fact that, in our context, the terms
$ \nabla_{\mu}\nabla_{\nu}\phi$ are asymptotically
sub-leading if compared to the Weyl tensor
 (recall that, for our solutions, $R_{\mu\nu} = - 
\nabla_{\mu}\nabla_{\nu}\phi$).

\renewcommand{\theequation}{5.\arabic{equation}}
\setcounter{equation}{0}
\section{Spherically symmetric little bangs?}

The solutions discussed in the previous sections
 have various symmetries: symmetries inherited from the 
 low energy effective action (like scale invariance and dilaton
 shift), accidental symmetries
(like parity) appearing in the solutions of the equations of motion and 
symmetries of the metric tensor. This last class of symmetries was imposed 
 from the very beginning \cite{m}, by requiring that our  geometry had
 two space-like killing vector fields  admitting an Abelian group
$G_2$ on space-like orbits $S_2$ \cite{kramer}. A legitimate question to ask 
is what does it change when we assume from the very beginning a different 
symmetry of the metric. Recent studies \cite{3} were dealing, for example,
with spherically symmetric manifolds arguing that the singularity is a
 (rather generic) property of string cosmology.
Moreover, those authors were more 
interested in the Einstein frame solutions that in the String frame ones 
(only for technical reasons).
 The question is then if it is possible 
to  find regular and geodetically complete
 solutions of the low energy beta functions in the Einstein 
frame with a spherically symmetric line element. This problem was 
recently discussed in the context of general relativity \cite{rev,c}
We present here a possible example in this direction.

We want to solve the low energy beta functions with a spherically
 symmetric ansatz and we want to see if they admit  regular solutions.
Suppose that the metric is spherically symmetric and with line element
\begin{equation} 
ds^2 =  f(r,t) \bigl[dt^2 - dr^2\bigr] -
 g(r,t)\bigl[ d\theta^2 + \sin^2{\theta} d\varphi^2\bigr],
\label{sph}
\end{equation}
Assume then, following the techniques exploited of Ref. \cite{m},
that $f(r,t)$ and $g(r,t)$ can be factorised as the product of two
regular functions only depending either on time or on space, namely
\begin{equation}
f(r,t) = e^{2 \mu r} \sigma(t),~~~g(r,t) = e^{ 2 \mu r} \rho(t)
\label{ans}
\end{equation}

Eqs. (\ref{beta1}) and (\ref{beta2}) can be easily written in 
the metric (\ref{sph}) with the ansatz (\ref{ans}). Since we want a
completely homogeneous  coupling we will set to zero  the spatial
derivatives of the dilaton. 
\begin{eqnarray}
& & 2 \ddot\phi - \frac{\dot\sigma}{\sigma} \dot\phi +
\biggl(\frac{\dot\rho}{\rho}\biggr)^2 + 
\biggl(\frac{\dot\rho}{\rho}\biggr) \biggl(\frac{\dot\sigma}{\sigma}\biggr)
+  \biggl(\frac{\dot\sigma}{\sigma}\biggr)^2 
- 2\frac{\ddot\rho}{\rho} - \frac{\ddot\sigma}{\sigma}
+ 4 \mu^2 =0~~~~~~~~~~~~~~~~~~~~~~~~~(00)
\label{00}\\
& &\dot\phi - \frac{\dot\sigma}{\sigma}
=0~~~~~~~~~~~~~~~~~~~~~~~~~~~~~~~~
~~~~~~~~~~~~~~~~~~~~~~~~~~~~~~~~~~~~~~~~~~~~~~~~~~~~~~(r0)
\label{r0}\\
& &\frac{\dot\sigma}{\sigma} \dot\phi
 - \frac{\dot\rho}{\rho} \frac{\dot\sigma}{\sigma} +
\biggl(\frac{\dot\sigma}{\sigma}\biggr)^2 - \frac{\ddot\sigma}{\sigma}
=0~~~~~~~~~~~~~~~~~~~~~~~~~~~~~~~~~~~~~~~~~~~~~~~~~~~~~~~~~~~~~~~~(rr)
\label{rr}\\
& & \frac{\ddot\rho}{\rho} + 2 \frac{\sigma}{\rho} -
\frac{\dot\rho}{\rho} \dot\phi - 4 \mu^2=0
~~~~~~~~~~~~~~~~~~~~~~~~~~~~~~~~~~~~~~~~~~~~~~~~~~~~~~~~~~~~~~~~~~
~~(\theta\theta)
\label{thth}\\
& &\frac{\dot\rho}{\rho} - \frac{\ddot\rho}{\rho} - 2 \frac{\sigma}{\rho}
+ 4 \mu^2 =0
~~~~~~~~~~~~~~~~~~~~~~~~~~~~~~~~~~~~~~~~~~~~~~~~~~~~~~~~~~~~~~~~~~~~~
(\varphi\varphi)
\label{phph}\\
& &2 \ddot\phi - \dot\phi^2 -\dot\phi \frac{\dot\rho}{\rho} 
- 2 \frac{\ddot\rho}{\rho} - 
\frac{\ddot\sigma}{\sigma} - 2 \frac{\sigma}{\rho} +
\biggl(\frac{\dot\rho}{\rho}\biggr)^2 
+2 \biggl(\frac{\dot\sigma}{\sigma}\biggr)^2 + 6 \mu^2= 0 
~~~~~~~~~~~~~~~~~~~~~~~(\phi)
\label{ph}
\end{eqnarray}
From Eq. (\ref{r0}) we get, integrating once, that, $\sigma =
\exp{[\phi]}$. The other equations give a consistent solution in the
form
\begin{equation}
\phi = 2 \mu t - \log{ (1 + 2 e^{2\mu t})} +\gamma,~~~~\rho= e^{2\mu t}
,~~~~\sigma=\frac{e^{2\mu t}}{2 e^{2 \mu t} + 1}
\label{solsph}
\end{equation}
notice that $\gamma$ is the integration constant associated with the
evolution of the dilaton; the over dot denotes the derivation with
respect to $t$. Since the action we are discussing is
invariant under dilaton shift, we can require that the dilaton starts
its evolution deeply in the perturbative regime where $\gamma$ will be
very negative and $g(\phi)\ll 1$.
The line element can then be written as
\begin{equation}
ds^2 = \frac{ e^{ 2 \mu (r + t)}}{ 2 e^{ 2 \mu t }+ 1} \bigl[ dt^2 -
dr^2\bigr] - e^{ 2\mu (r + t)} \bigl[d\theta^2 + 
\sin^2{\theta} d\varphi^2\bigr].
\label{metsp}
\end{equation}
 We would be tempted to spot this as an example of a (spherically
 symmetric) little bang dynamics. This does not seem the case.
In fact there are quite amusing features associated with this solution which
makes it in a sense unphysical but also interesting, in another sense.
The curvature invariants associated with this solution are all
regular (see Appendix C for the results). In order to understand why
this example is qualitatively different from the ones discussed in the
previous sections we can look at the asymptotic behaviour of the
dilaton coupling:
\begin{equation}
\lim_{t\rightarrow +\infty }\phi(t) = -\sqrt{2},~~~
\lim_{t\rightarrow -\infty }\phi(t) = 2 \mu t.
\end{equation}
Thus $\dot\phi$ goes to zero for $t\rightarrow +\infty$ and it goes to
a constant (i.e. $\dot\phi \sim 2\mu$) for $t\rightarrow - \infty$.
This behaviour has to be contrasted with the one exhibited by the
solutions reported in Sec. 1 where $\dot\phi$ was going to zero in
both limits. From a physical point of view we would prefer a coupling
evolving between two asymptotically constant regimes. However one might
speculate that a linearly evolving dilaton might be, after all, not so
wrong especially in connection with the analysis of initial conditions
in the pre-big-bang scenario \cite{3,6}. The second nasty feature of
the solution (see Appendix C for details) is that the curvature
invariants are regular in space but they exponentially 
diverge in time (in the String frame) for $t\rightarrow -\infty$.
This is clearly a fatal feature for this solution. It is however
interesting to notice that the curvature invariants are regular and
well defined in the Einstein frame. The reason for this drastically
different behaviour in the two frames is the linear evolution of the
dilaton (for $t\rightarrow -\infty$) which appears in the conformal
factor defining the relation between the Einstein and the String
frame. Unfortunately, looking at the Einstein frame picture, we cannot
``see'' the dynamics of a ``little bang'' 
since the curvature does not grow to a
maximal value. On the other hand
we see from the analytical form of the curvature invariants (Appendix
C) that the curvature scalar decreases from a constant value.

From this example we draw two important conclusions. First of
all, when dealing with an initial state with linear dilaton (as in
Ref. \cite{3}) regular solutions in the Einstein frame turn singular
in the String frame. This does not exclude the interesting possibility that
singular solutions (in the Einstein frame) can turn into spherically
symmetric (regular) solutions in the String frame.

The second thing we learn from this example is that, apparently, the
possibility of having a growing dilaton evolving from a constant value
to another constant value (in such a way that the curvature invariants
will be bounded and regular in both frames) seems to be associated to
the cylindrical symmetry of the metric. We cannot say if this is a
mere technical point (which helps in finding an analytical solution)
or if it is somehow deeper than that. We hope to come back on this
point in the ner future.
 
We want finally to recall that 
the solution (\ref{solsph}) is nothing but the string cosmological 
version of the one discussed in \cite{rev,c} (notice that in
Ref. \cite{c} other examples of homothetic cosmological space-times
are given whereas in \cite{rev} the geodesics of this model are
discussed). 

\renewcommand{\theequation}{6.\arabic{equation}}
\setcounter{equation}{0}
\section{ Conclusions and speculations}

In this paper we discussed the singularity properties of a class of
solutions of  the low energy beta functions. The space-time we
described turns out to be singularity free in the technical sense
since it has the curvature invariants all bounded and since it is
also geodetically complete. We discussed the peculiar features of our
space-time in the light of the singularity theorems and we saw that
at least one of their hypotheses was violated. In particular we found
examples where the only condition to be violated is the existence of
trapped (closed) surfaces. We also discussed the physical
interpretation of this class of models. They describe, in a string
cosmological context, the occurrence of an inhomogeneous phase with
growing curvature and coupling. The dynamics of the model is driven
by the kinetic energy of the dilaton coupling and the big-bang
singularity is replaced by a maximal curvature phase which we named 
little bang. The bang is associated with the maximal curvature which
can be as large as the string curvature scale. The bang 
is small  in the sense that the solutions never develop  a
singularity. Since the
maximal curvature scale is always smaller than one (in string units)
the tree level action is appropriate for the description of such a
phase.
These analytical examples show that it possible to regularise the
singularities appearing in the low energy beta  functions with strong
inhomogeneities. Our model describes, then, what we called
inhomogeneous bangs. These example also show that it is not a must to
include corrections in the string tension expansion in order to have
regular curvature invariants. We stress that , in our case, the
geodesics are also complete and, therefore, no singularity appears in
the geodesics for any finite value of the affine parameter. Since our
solutions are defined from $t\rightarrow -\infty$ to $t\rightarrow
+\infty$ and since,  for negative times, the dilaton coupling  and
the curvature are always increasing, it seems natural to find a
contact between the models of inhomogeneous bang and the pre-big-bang
models. Apparently the similarities are quite deep starting from the
fact that the equations of motion of the fields in these two models
seem very similar.

We also discussed some spherically symmetric examples whose
physical properties seem quite intriguing but, in our opinion, still 
not completely physical. 

Finally we stress that our solutions are always
inhomogeneous. The physical picture we have in mind is therefore the
following. We would like to find a model where a phase of
(homogeneous) growing dilaton coupling is replaced around $[t,x]=0$
by an inhomogeneous phase. No explicit models of this kind were
constructed, up to now. At the same time recent investigations
\cite{int} showed that the occurrence of these type of models is not
excluded (in a general relativity context) since explicit general
families of solutions were constructed where the ordinary
Friedmann-Robertson-Walker models appear together with other
singularity free (inhomogeneous) perfect fluid models. 
Our context is certainly
different, but, none the less,  we think that it might be useful to
investigate the possible occurrence of such a scenario in a
superstring cosmological framework. 

\section*{Acknowledgments}

I would like to thank J. M. M.  Senovilla for very useful communications and
 G. Veneziano for kind interest in this investigation.

\newpage

\begin{appendix}
\section*{APPENDIX}
\renewcommand{\theequation}{A.\arabic{equation}}
\setcounter{equation}{0}
\section{Christoffel Symbols}
We report, in this appendix the Christoffel symbols computed in the
case of the metric given in Eq. (\ref{metric}). In Table \ref{tab1}
we report the general expression of the Christoffel connections in
the case of a metric with two Abelian killing vectors which are
hypersurface orthogonal and orthogonal to each others. From the
expressions given in Table \ref{tab1}, using the explicit form of our
solutions given in Eq. (\ref{metric}) we get, after some simple
algebra,
\begin{table}
\begin{center}
\begin{tabular}{|c|c|}
\hline
 $\Gamma_{x x}^{x} = \frac{1}{2}\frac{\partial \log{A}}{\partial x}$
&
 $\Gamma_{y y}^{x}  =
 -\frac{C~B}{2~A}\frac{\partial \log{(B~C)}}{\partial x}  $ \\
\hline
 $ \Gamma_{zz}^{x} = - \frac{B}{2~ A ~C}
\frac{\partial\log{(B~C)}}{\partial x}
 $ & $\Gamma_{x t}^{x} =
\frac{1}{2}\frac{\partial\log{A}}{\partial~t} $\\
\hline
 $\Gamma_{t t}^{x} =\frac{1}{2}\frac{\partial \log{A}}{\partial x}$ &
$\Gamma_{x y}^{y}=
 \frac{1}{2}\frac{\log{(B~C)}}{\partial x}$ \\
\hline
 $\Gamma_{x t}^{y} = \frac{1}{2}\frac{\log{(B~C)}}{\partial t} $ &
$\Gamma_{x z}^{z}=
 \frac{1}{2}\frac{\partial\log{(B/C)}}{\partial x} $ \\
\hline
 $  \Gamma_{zt}^{z}=\frac{1}{2}\frac{\partial\log{(B/C)}}{\partial t}
 $
& $\Gamma_{x x}^{t}=\frac{1}{2} \frac{\partial\log{A}}{\partial t}$
\\
\hline
$ \Gamma_{y y}^{t}=
\frac{C~B}{2~A}\frac{\partial\log{(C~B)}}{\partial t}$ &
$ \Gamma_{zz}^{t}=\frac{B}{2~A~C} \frac{\partial\log{(B/C)}}{\partial
t}$\\
\hline
$\Gamma_{x t}^{t} = \frac{1}{2} \frac{\partial\log{A}}{\partial x}$&
$
\Gamma_{t t}^{t}
=\frac{1}{2}\frac{\partial\log{A}}{\partial t}$\\
\hline
\end{tabular}
\end{center}
\caption[a]{We report the Christoffel symbols computed in a geometry
with two Abelian killing vector fields which are hyper-surface
orthogonal and orthogonal to each others. The notations are the ones
used in Section II.}
\label{tab1}
\end{table}

\begin{eqnarray}
&&\Gamma^{t}_{t t} = \frac{\mu}{2} \frac{ \alpha + (\beta +
2)\sinh{\mu t}}{\cosh{\mu t}}, ~~~ \Gamma^{t}_{xx} = \Gamma^{t}_{t
t},
\nonumber\\
&&\Gamma^{t}_{x t} = \frac{\mu}{2} \beta (\beta +
1)\tanh{\biggl(\frac{\mu x}{2}\biggr)},~~~\Gamma^{t}_{t x} =
\Gamma^{t}_{x t},
\nonumber\\
&&\Gamma^{t}_{y y} = \mu\frac{\bigl[ \alpha + ( \beta + 2) \sinh{\mu
t}\bigr]}{ \bigl[ \sinh{\bigl(\frac{\mu x}{2}\bigr)}\bigr]^{-2}
\bigl[\cosh{\bigl(\frac{\mu x}{2}\bigr)}\bigr]^{ 2 \beta^2 - 2}
\sinh{\mu t}},
\nonumber\\
&&\Gamma^{t}_{zz} =\mu \frac{ \alpha - \beta \sinh{\mu t}}{
\bigl[\cosh{\bigl(\frac{\mu x}{2}\bigr)}\bigl]^{ 4 \beta + 2 \beta^2}
\bigl[\cosh{\mu t}\bigr]^{3 + 2 \beta}},
\nonumber\\
&&\Gamma^{x}_{x x}= \frac{\mu}{2} \beta(\beta + 1)
\tanh{\biggl(\frac{\mu x}{2}\biggr)},~~~\Gamma^{x}_{t t}=
\Gamma^{x}_{x x}
\nonumber\\
&&\Gamma^{x}_{x t} = \frac{\mu}{2} \frac{ \alpha + (\beta +
2)\sinh{\mu t}}{\cosh{\mu t}}, ~~~ \Gamma^{r}_{t r} = \Gamma^{r}_{r
t},
\nonumber\\
&&\Gamma^{x}_{y y} =\frac{\mu}{2} \biggl[\sinh{\biggl(\frac{\mu
x}{2}\biggr)} \biggr]\frac{\bigl[\beta - (\beta +2) \cosh{\mu
t}\bigr]}{\bigl[\cosh{\bigl(\frac{\mu x}{2}\bigr)}\bigr]^{ 2 \beta^2
-1}}
\nonumber\\
&&\Gamma^{x}_{z z} = \mu \beta \frac{ \tanh{\bigl( \frac{\mu
x}{2}\bigr)}}{\bigl[ \cosh{\bigl( \frac{\mu x}{2}\bigr)}\bigr]^{ 4
\beta + 2 \beta^2 }\bigl[\cosh{\mu t}\bigr]^{2 (\beta + 1)}},
\nonumber\\
&&\Gamma^{y}_{y x} = \frac{\mu }{2} \frac{(\beta + 2) \cosh{\mu x} -
\beta }{\sinh{\mu x}},~~~\Gamma^{y}_{x y} = \Gamma^{y}_{y x},
\nonumber\\
&&\Gamma^{y}_{y t} = \frac{\mu}{2} \frac{ \alpha + (\beta + 2)
\sinh{\mu t}}{\cosh{\mu t}},~~~\Gamma^{y}_{t y} = \Gamma^{y}_{y t},
\nonumber\\
&&\Gamma^{z}_{x z}= -\frac{\mu}{2}\beta \tanh{\biggl(\frac{\mu
x}{2}\biggr)},~~~\Gamma^{z}_{z r} = \Gamma^{z}_{x z},
\nonumber\\
&&\Gamma^{z}_{z t} = \frac{\mu}{2} \frac{\alpha - \beta\sinh{\mu
t}}{\cosh{\mu t}},~~~\Gamma^{z}_{t z} = \Gamma^{z}_{z t}.
\end{eqnarray}

{}From Eq. (\ref{metric}) the general form of the geodesics can be
derived
\begin{eqnarray}
&&\frac{d^2 t}{d \lambda^2} + \frac{1}{2} \frac{\dot{A}}{A} \biggl[
\biggl(\frac{d t}{d\lambda}\biggr)^2
+ \biggl(\frac{d x}{d \lambda}\biggr)^2\biggr] + \frac{A'}{A} \frac{d
t}{d \lambda}\frac{d x}{d\lambda}  + \frac{ \dot{B}C - \dot{C}B}{ 2 A
C} \biggl(\frac{d z}{d \lambda}\biggr)^2 + \frac{ \dot{B} C + \dot{C}
B}{2 A} \biggl(\frac{dy}{d\lambda}\biggr)^2=0
\nonumber\\
&& \frac{ d^2 x}{d \lambda^2}  + \frac{1}{2}\frac{A'}{A}
\biggl[\Biggl(\frac{d t}{d \lambda}\biggr)^2 +\biggl(\frac{d x}{d
\lambda}\biggr)^2 \biggr]+\frac{\dot{A}}{A} \frac{d t}{d \lambda}
\frac{dx }{d\lambda} - \frac{ B'C + C' B}{2 A
C^2}\biggl(\frac{dz}{d\lambda}\biggr)^2 - \frac{ B' C +  C' B}{2 A}
\biggl(\frac{dy}{d \lambda}\biggl)^2=0
\nonumber\\
&&\frac{ d^2 y}{d \lambda^2} + \biggl(\frac{\dot{B}}{B}
+\frac{\dot{C}}{C}\biggr) \frac{d t}{d \lambda} \frac{ d x}{d
\lambda} + \biggl( \frac{B'}{B} + \frac{C'}{C}\biggr) \frac{d x}{d
\lambda} \frac{ d y}{d \lambda} =0
\nonumber\\
&&
\frac{ d^2 z}{ d\lambda} + \biggl(\frac{ \dot{B}}{B} -
\frac{\dot{C}}{C} \biggr) \frac{d z}{d \lambda}  \frac{ d
t}{d\lambda} + \biggl( \frac{B'}{B} - \frac{C'}{C}\biggr) \frac{ d
z}{ d \lambda} \frac{d x}{d \lambda}=0
\nonumber\\
&&\biggl(\frac{d t}{d\lambda}\biggr)^2 - \biggl(\frac{d
x}{d\lambda}\biggr)^2 = \frac{ B C }{A} \biggl( \frac{d
y}{d\lambda}\biggr)^2 +\frac{ B }{C A} \biggl(\frac{ d z}{d\lambda}
\biggr)^2 +\frac{ {\bf s}}{A}
\end{eqnarray}
Notice that (only in this formula) we denoted with an over dot the
derivation with respect to $t$ and with a prime the derivation with
respect to $x$. In the rest of the paper the prime denotes simply the
derivation with respect to the affine parameter $\lambda$. We
apologise with the reader for the possible confusion.

\renewcommand{\theequation}{B.\arabic{equation}}
\setcounter{equation}{0}
\section{Geodesic Convergence}

 The geodesic convergence condition tells us that, given any
non-space-like vector $u$ we have that
\begin{equation}
u^{\mu} u^{\nu} R_{\mu\nu} >0.
\label{a}
\end{equation}
 We will show that this condition is very important and interesting
for the class of solutions we are discussing. In fact we can show
that not for all choices of the  parameter $\beta$ this condition is
satisfied. In order to show this let us report the components of the
Ricci tensor
\begin{eqnarray}
&&R_{x}^{x} = -\frac{ \alpha \mu ^2}{2}\frac{\bigl[ \alpha + 2
\sinh{\mu t} + \beta \sinh{\mu t}
\bigr]}{\biggr[\cosh{\biggl(\frac{\mu x}{2}\biggr)}\biggl]^{2 \beta
(\beta + 1)} \biggl[ \cosh{\mu t}\biggr]^{2 +\beta}} e^{-\alpha
gd(\mu t)}
\nonumber\\
&&R_{y}^{y}= -\frac{\alpha \mu^2}{2} \frac{\bigl[ \alpha + \beta
\sinh{\mu t}\bigr]}{\biggl[\cosh{\biggl(\frac{\mu
x}{2}\biggr)}\biggr]^{2\beta(\beta + 1)} \bigl[\cosh{\mu t}\bigr]^{4
+ \beta}} e^{- \alpha gd(\mu t)}
\nonumber\\
&&R_{z}^{z} = -\frac{\alpha \mu^2}{2} \frac{\bigl[ \alpha - \beta
\sinh{\mu t}\bigr]}{\biggl[\cosh{\biggl(\frac{\mu
x}{2}\biggr)}\biggr]^{2\beta (\beta + 1)} \bigl[ \cosh{\mu t}\bigr]^{4
+ \beta}} e^{- \alpha gd(\mu t)}
\nonumber\\
&&R_{x}^{0} = \frac{\alpha \beta( 1 +\beta)\mu^2
}{2}\biggl[\sinh{\biggl(\frac{\mu
x}{2}\biggr)}\biggr]\biggl[\cosh{\biggl(\frac{\mu
x}{2}\biggr)}\biggr]^{-1 - 2\beta (1 +\beta)} \bigl[ \cosh{\mu
t}\bigr]^{ -3 -\beta} e^{-\alpha gd(\mu t)}
\nonumber\\
&&R_{0}^{0} = \frac{\mu^2}{2} \frac{\bigl[ \beta^2 -
4 + \alpha(4+ \beta)\sinh{\mu t}\bigr]}{\bigl[\cosh{\mu t}\bigr]^{4 +
\beta}\biggl[\cosh{\biggl(\frac{\mu
x}{2}\biggr)}\biggr]^{2\beta(\beta + 1)} }e^{-\alpha gd(\mu t)}
\end{eqnarray}
Notice that something very amusing happens from this formulas. We
point out that our class of exact solutions contains particular
solutions where the geodesic convergence condition is clearly
satisfied. Take for example $\alpha= 2\sqrt{3}$, then, since $\beta =
\pm \sqrt{\alpha^2 + 4}$ we have that, for the value $\beta = -
\sqrt{\alpha^2+ 4}$ $u^{\mu} u^{\nu} R_{\mu \nu}>0$. Of course in the
opposite case (i.e. $\beta = 4$) there is no definite sign for
$u^{\mu} u^{\nu} R_{\mu\nu}$. As a consequence, since the geodesic
convergence condition changes sign as an hyperbolic cosine the
geodesic convergence will not occur.

It is interesting to notice that the geodesic convergence condition is always 
completely satisfied when translated to the Einstein frame where the low 
energy beta functions can also be written and solved \cite{m}. 

The Einstein frame is defined by the conformal transformation which
 diagonalises the gravitational action by decoupling the dilaton from the 
Einstein-Hilbert Lagrangian. In the Einstein frame the Planck scale ( 
$\lambda_{P}\sim 10^{-33}~{\rm cm}$ is constant whereas the string scale 
evolves according to $\lambda_{s}(t)= e^{- \phi/2} \lambda_{P}$.  Since (in 
four dimensions) the Einstein and String frame dilatons are exactly equal, 
the string ($g_{\mu\nu}$) and Einstein ($G_{\mu\nu}$) metrics are connected by
the obvious conformal rescaling $g_{\mu\nu}= e^{\phi} G_{\mu \nu}$. Notice 
that this conformal rescaling leads to changes in the geodesic convergence 
condition given in Eq. (\ref{a}). In fact in the Einstein frame  we obtain that
\begin{equation}
{\cal R}_{0}^{0} =  \frac{\beta^2 -4}{2} 
\mu^2\bigl[ \cosh{\mu t}\bigr]^{-4 - \beta} 
\biggl[ \cosh{\biggl(\frac{\mu x}{2}\biggr)}\biggr]^{-2 \beta( \beta + 1)}
\end{equation}
(${\cal R}_{\mu}^{\nu}$ is the Ricci tensor in the Einstein frame). From this
 last equation we conclude that the geodesic convergence 
condition is satisfied without restrictions (in the Einstein
 frame) since the physical range of $\beta$ is $|\beta|>2$. This 
simple exercise can also show that the formulation of the singularity theorems 
in the String and  Einstein frames can be different.

\renewcommand{\theequation}{C.\arabic{equation}}
\setcounter{equation}{0}
\section{A spherically symmetric example}

In this section of the Appendix we report the curvature invariants
associated with the spherically symmetric solution of the low energy
beta functions discussed in Sec. 5. In the String frame the curvature
invariants, computed from Eq. (\ref{metsp}) are:
\begin{eqnarray}
&&R^{\mu\nu} R_{\mu\nu} = 2\mu^4 \frac{ 1 + 4 e^{2 \mu t} + 20 e^{4 \mu
t}}{\bigl[1 + 2 e^{ 2 \mu t}\bigr]^2}e^{ - 4 \mu ( r + t)},
\\
&&R^{\mu\nu\alpha\beta}R_{\mu\nu\alpha\beta} = 2\mu^4 \frac{ 1 + 4 e^{2
\mu t} + 20 e^{4 \mu t}}{\bigl[1 + 2 e^{ 2 \mu t}\bigr]^2}
e^{ - 4 \mu ( r + t)},
\\
&&C^{\mu\nu\alpha\beta} C_{\mu\nu\alpha\beta}= \frac{4 \mu^4}{3} \frac{
\bigl[ 2 e^{ 2 \mu t} -1\bigr]^2}{\bigl[ 2 e^{ 2 \mu t} +
1\bigr]^2}e^{ - 4\mu (r + t)},
\\
&&R^2 =\frac{4 \mu^4}{3} \frac{
\bigl[ 2 e^{ 2 \mu t} -1\bigr]^2}{\bigl[ 2 e^{ 2 \mu t} +
1\bigr]^2}e^{ - 4\mu (r + t)}
\end{eqnarray}
As we can see the curvature invariants explode exponentially for
$t\rightarrow -\infty$.
It is instructive to compute the curvature invariants also in the
Einstein frame metric. They are:
\begin{eqnarray}
{\cal R}^2 &=& \frac{4\mu^4}{ e^{ 4\mu r}~ \bigl[ 1 + 2 e^{ 2\mu t}\bigr]^4},
\nonumber\\
{\cal R}_{\mu\nu} ~{\cal R}^{\mu\nu} &=& \frac{4\mu^4}{e^{4\mu r} \bigl[ 1 + 2
e^{2 \mu t}\bigr]^4},
\nonumber\\
{\cal R}_{\mu\nu\alpha\beta} {\cal R}^{\mu\nu\alpha\beta} &=&
\frac{12\mu^4 }{ e^{4\mu r} ~\bigl[ 1 + 2 e^{2\mu t}\bigr]^4},
\nonumber\\
{\cal C}_{\mu\nu\alpha\beta} {\cal C}^{\mu\nu\alpha\beta} &=&
\frac{16\mu^4}{ 3~e^{4\mu r}~ \bigl[ 1 + 2 e^{2\mu t}\bigr]^4},
\end{eqnarray}
We clearly see from these equations that the curvature invariants are
regular.  However, something unpleasant happens: for $t\rightarrow
-\infty$ the invariants do not tend to zero but to a constant. Notice
that one of the main point of the examples discussed in \cite{m} was
that {\em both} in the Einstein and in the String frame the curvature
invariants were going to zero for $t\rightarrow \pm \infty$. The
peculiar behaviour of our spherically symmetric example seems not
completely physical and impossible to associate with a little ban in
the sense discussed in this paper.

\end{appendix}

\end{document}